
\input phyzzx
\date={October 1992}    
\Pubnum={\caps UPR--534--T}
hep-th/9210123

\def\to{\rightarrow}

\titlepage
\title{ Cauchy Horizons, Thermodynamics and Closed Time-like Curves in
Planar Supersymmetric
Space-times}

\frontpageskip=0.5\medskipamount plus 0.5 fil
\author{ Mirjam Cveti\v c\foot{email CVETIC@cvetic.hep.upenn.edu}
Richard Davis\foot{email DAVIS@ovrut.hep.upenn.edu} \break
Stephen Griffies\foot{email GRIFFIES@cvetic.hep.upenn.edu}
and Harald H. Soleng\foot{email SOLENG@steinhardt.hep.upenn.edu or
               SOLENG@vuoep6.uio.no}
\foot{On leave from the University of Oslo, Norway} }
\address{Department of Physics\break
        University of Pennsylvania\break
        209 So. 33rd Street \break
        Philadelphia, PA 19104--6396}

\abstract{We study geodesically
complete, singularity free space-times   induced
by supersymmetric planar domain walls interpolating between
Minkowski and anti-de Sitter ($AdS_4$) vacua.
A geodesically complete space-time without
closed time-like curves includes
an infinite number of semi-infinite Minkowski space-times,
separated from each other by a region of $AdS_4$ space-time.
These space-times are closely  related to the
extreme Reissner Nordstr\" om (RN) black hole, exhibiting Cauchy
horizons with
zero  Hawking
temperature, but in contrast to the RN black hole there is no entropy.
Another geodesically complete extension with closed time-like curves
involves  space-times connecting
a finite number of semi-infinite Minkowski space-times.}

PACS numbers 04.20.-q, 04.65.+e, 11.30.Pb, 12.10.Gq

\endpage

\unnumberedchapters

\REF{\CGRI}{ M. Cveti\v c, S. Griffies, and S.-J. Rey, Nucl. Phys.
\bf B381, \rm 301 (1992).}

\REF{\CG}{M. Cveti\v c and  S. Griffies, Phys. Lett. \bf 285B, \rm 27 (1992).}

\REF{\GHW}{See for example G.W. Gibbons, C.M. Hull, N.P. Warner,
Nucl. Phys. \bf B218, \rm 173 (1983);
C.M Hull, Nucl. Phys \bf B239, \rm 541 (1984) and references therein.}

\REF{\CGRII}{ M. Cveti\v c, S. Griffies, and S.-J. Rey, \sl Nonperturbative
Stability of Supergravity and Superstring Vacua, \rm
UPR-494-T, hep-th/9206004
(May 1992), Nucl. Phys. {\bf B}, in press.}

\REF{\CD}{S. Coleman and F. De Luccia,
Phys. Rev. \bf D21 \rm , 3305 (1980).}

\REF{\CQR}{M. Cveti\v c, F. Quevedo and S.-J. Rey,
Phys. Rev. Lett. \bf 63, \rm 1836 (1991).}

\REF{\AT}{ E. Abraham and P. Townsend, Nucl. Phys. \bf 351B, \rm 313 (1991).}

In this Letter we analyze  geodesically complete space-times induced by
supersymmetric domain walls interpolating between
$3+1$ Minkowski and Anti-deSitter
($AdS_4$) vacua.\refmark{\CGRI\ , \CG}
These are
planar and causally
non-trivial exact solutions to Einstein's equations
without space-time singularities.
In addition, they provide a new class of
space-times, asymptotically Minkowski in a direction
away from the wall, in which the
thermodynamic properties of Cauchy horizons and the physics of
closed time-like curves can be studied, and
with the attractive feature that the curvature is everywhere finite.

The ADM mass of  supersymmetric (Minkowski  or $AdS_4$)
vacua in $N=1$, $d=4$ supergravity
vanishes\refmark{\GHW}, thus ensuring the degeneracy
of supergravity vacua.  A complementary way  of establishing
stability of such vacua exploits
\refmark{\CGRII}
the existence  of a
Bogomol'nyi bound  for
ADM mass/area stored in the
wall of the bubble  tunnelling
between two supersymmetric vacua.
Namely, the minimal ADM mass/area of such a bubble wall
is compatible with the Coleman-DeLuccia bound\refmark{\CD}
only in the  limit of  infinite bubble radius,\ie , in
the limit when the tunnelling is absolutely supressed.
As a consequence of vacuum degeneracy
one expects domain wall solutions
interpolating between isolated supersymmetric
vacua not only in the global case\refmark{\CQR\ ,\AT},
but also when gravity is turned on.

Domain walls interpolating between
non-degenerate supersymmetric minima of a supergravity matter
potential were found in Ref.\refmark{\CGRI} and further
studied in Ref.\refmark{\CG}.
A field theoretical realization
of such walls exists within
$N=1, d=4$ supergravity coupled to chiral superfields.
We consider the simplest case of
one chiral supermultiplet, ${\cal T}$,
with a nonzero superpotential,
$W({\cal T})$. The bosonic
part of the Lagrangian density (consisting of the metric $g_{\mu \nu}$
and the complex scalar field component
$T$) is
$$
L = -{1 \over 2 \kappa}R + K_{T \bar T}g^{\mu \nu}\partial_{\mu}\bar T
\partial_{\nu}T - e^{\kappa K}(K^{T \bar T}|D_{T}W|^{2} - 3 \kappa |W|^{2})
\eqn\lagrangian$$
where
$K(T, \bar T) =$ K\"ahler potential,  $D_{T}W = e^{-\kappa K} (\partial_T
e^{\kappa K} W)$ and $\kappa = 8\pi G$.
Supersymmetric vacua  satisfy  $D_TW=0$; therefore, supersymmetric
vacua with $W=0$ correspond to Minkowski space-times
while those with $W\neq 0$ correspond to
$AdS_{4}$ space-times with cosmological constant
$\Lambda = -3|\kappa W e^{{\kappa K \over 2}}|^{2} =
 -3\alpha^{2}$.

One of the minimal energy domain wall
solutions corresponds to static, planar
domain walls\refmark{\CGRI ,\CG}
interpolating  between
an
isolated supersymmetric
Minkowski vacuum
as $z\to\infty$  and
an
isolated supersymmetric
$AdS_4$ vacuum  as
$z\to -\infty$. The matter
and the metric coefficients satisfy the first order
differential equations
corresponding to the supersymmetric bosonic backgrounds. One finds
a  static  solution  both
for the scalar  matter field $T(z)$ and the conformally  flat
metric:\refmark{\CGRI ,\CG}
$$
ds^{2} = A(z)(dt^{2} - dx^{2} - dy^{2} - dz^{2}).
\eqn\metric$$
The conformal factor $A(z)$ has the asymptotic  behaviour:
$$
A(z) \to \left\{
\eqalign{
1, \ \   &z \to +\infty  \cr
(\alpha z)^{-2}, \ \ &z \to -\infty }
\right.
\eqn\sol$$
while in the wall region ($z\sim 0$) it smoothly
interpolates between the  two regions.
Here, $\Lambda = -3\alpha^{2}$ is the
cosmological constant  of  the $AdS_4$ vacuum.
The wall's
ADM mass/area
$\sigma = (2/\sqrt{3}) \kappa^{-1}|\Lambda|^{1/2} \equiv 2\alpha$
saturates the Bogomol'nyi bound.

\REF{\ISR}{W. Israel, Nuovo Cimento \bf 44B, \rm  1 (1966).}

\REF{\TOLMAN}{R. C. Tolman, Phys. Rev. \bf 35, \rm 875 (1930).}


The possibility of a static
juxtaposition of $AdS_{4}$ and Minkowski space-times
can be understood from a general relativistic perspective
without referring  to the underlying matter field configuration.
For this purpose we approximate the wall
as infinitely thin
with the conformal
factor $A(z)$  changing from the $AdS_{4}$ form, $A(z)=
(\alpha z)^{-2}$,
to the Minkowski
form, $A(z)=1$, at a chosen value of $z=z_{0} \equiv -\alpha ^{-1}$.
Because Einstein's tensor is of second order in
derivatives of the metric,
it has a $\delta$-function singularity at $z_0$, 
and we therefore use
Israel's \refmark{\ISR} method of  singular
hypersurfaces. The wall is
found to satisfy the equation of state of a domain wall.
By adapting Tolman's
\refmark{\TOLMAN} mass formula to
give a gravitational surface mass density,
the effective gravitational mass/area  of the wall
is calculated to be $-2\alpha$
whereas that of the semi-infinite $AdS_{4}$
space-time is $2\alpha$.  The exact cancellation of the effective
masses  allows for a semi-infinite Minkowski space-time of zero
effective gravitational mass to exist adjacent to
the wall with the semi-infinite $AdS_{4}$ space-time on the other side.
The above result follows from Einstein's equations and the
assumed form of the metric.
Supergravity provides a
field theoretic realization of such  domain walls
with finite thickness and an ADM mass/area
precisely cancelling the effective gravity of the $AdS_{4}$ vacuum.



\REF{\NOTEVIII}{The spherical coordinates for
pure $AdS_{4}:$
$ds^2 = (\alpha \cos\psi)^{-2}
(dt_{c}^{2} - d\psi^{2} - \sin^{2}\psi d\Omega_{2}^{2})$
where
$-\pi \le t_{c} \le \pi, 0 \le \psi < \pi/2$,
are related to the planar coordinates through
$\alpha t = -{\cos t_{c} \over \sin t_{c} - \cos\theta \sin\psi} \   \
\alpha z = {\cos \psi  \over \sin t_{c} - \cos\theta \sin\psi} \   \
\alpha x = {\sin\psi \sin\theta \cos\phi
             \over \sin t_{c} - \cos\theta \sin\psi} \   \
\alpha y = {\sin\psi \sin\theta \sin\phi
             \over \sin t_{c} - \cos\theta \sin\psi}$
yielding $ds^{2} = (\alpha z)^{-2}(dt^{2} - dx^{2} - dy^{2} - dz^{2})$.
The transformation is defined on a particular
null diamond, for example, $0 \le t_{c} \pm \psi \le \pi$, where
time-like geodesics live.
Under this transformation, the transverse time-like geodesics
$z^{2} - t^{2} = (\alpha \epsilon)^{-2}$ become the periodic
radial geodesics
$\sin^{2}\psi =
[ (1-(\alpha \epsilon)^{-2} ) / ( 1+ (\alpha \epsilon)^{-2} ) ]^{2}
\sin^{2}t_{c}$.
The planar coordinates
with $z<0$ combined with $z>0$ completely cover all of
$AdS_{4}$\refmark{\CG}.}

\REF{\RINDLER}{ See for example W. Rindler, \sl Essential Relativity \rm,
Springer-Verlag 1979. }

We now discuss the space-time induced by the metric  \metric , \sol \ .
We first study the motion of test particles
which define the geodesics\refmark{\CG}.
The nulls are trivial since the metric is conformally flat.
In addition, due to  the boost invariance of the metric
in the  $x,y$ directions, we can without loss of generality move
to a frame in which a test particle moves only
transverse to the wall. Thus, it is sufficient to study
time-like geodesics in the $1+1$ system
$ds^{2} =
A(z)(dt^{2} - dz^{2})$.
As the metric is static, there is a
conserved energy parameter $\epsilon = A(z) dt/d\tau$ where $\tau$
is the proper time.
For time-like world-lines on the $AdS_{4}$ side (where $A(z) = (\alpha
z)^{-2}$)
the equation of motion yields\refmark{\CG}
$$z^{2} - t^{2} = (\alpha \epsilon)^{-2}.
\eqn\rindlerpath$$
This is the same world-line as
a particle undergoing
a constant proper acceleration of $(\alpha \epsilon)$
moving in a Minkowski space-time, \ie ,  a Rindler particle\refmark{\RINDLER}.
The constant scalar curvature of $AdS$
yields, through the equivalence principle, a
freely falling Rindler particle\refmark{\NOTEVIII}.

\REF{\FOOTX}{In Refs. \refmark{\CGRI} and \refmark{\CG} the geodesic
incompleteness of the $AdS_4$ space-time induced by the
domain walls was not recognized.}

The proper distance of a constant time slice, $d(z)
= \int^{z} \sqrt{A(z')}dz'$ is logarithmically
divergent on the $AdS_{4}$ side.
However, the time-like geodesics leaving the
wall at $\alpha z = -1$ and moving into the $AdS_{4}$ side
reach
$z=-\infty$
with $t = \infty$ in the finite
proper time $\tau =
-\int_{-1/(\alpha \epsilon)}
^{-\infty} A(z)dz/\sqrt{ \epsilon^2 -  A(z)}
\approx  \pi/2\alpha,$  where $A(z) \approx (\alpha z)^{-2}$ was used.
Therefore, the
coordinates $z,t,$ which completely cover the semi-infinite
Minkowski side, are {\it not} geodesically complete on the $AdS_{4}$
side\refmark{\FOOTX}.
To make the space-time geodesically complete
we must
extend the $AdS_4$ side
beyond the horizon at $z=-\infty$ onto a new patch.
On this new patch, we define a
new
coordinate $z'$ and identify
$z'=+\infty$ with the
old coordinate $z$ at $z=-\infty$.
In  addition, $t=\infty$ becomes $t'=-\infty$
upon crossing the horizon,
which continues the forward flow in the time-like direction.
By introducing
the new semi-infinite $AdS_{4}$
region, we have added $2\alpha$ to the effective gravitational
mass/area  of the system, and to remain in gravitational
equilibrium we
place an identical domain wall centered at $z'_0=+\alpha^{-1}$
The smooth extension of the scalar field creating the new wall is
$T(z') = T(-z)$.
This new
wall interpolates between the new $AdS_{4}$
region and another semi-infinite Minkowski
space-time.

\REF\FOOTIV{There are analogous
geodesically complete
space-times  for
domain walls interpolating between two
supersymmetric $AdS_{4}$ vacua, classified as type II and type
III domain walls in Ref. \refmark\CG . For type
II walls ($AdS_{4}-AdS_{4}$
walls
with superpotential
$W$ passing through $W=0$)
the Penrose conformal diagram (for a smooth
extension of the scalar field across the horizon) fills out the plane.
For  type III  walls ($AdS_{4}-AdS_{4}$ walls where
$W$ does not pass through $W=0$)
the space-time essentially corresponds to the
$AdS_{4}$ with a $z$ dependent
cosmological constant.
In this case the geodesically  complete Penrose
diagram
(again for a
smooth extension of the scalar field)
is a strip, \ie , the  middle of the $AdS_{4}$-Minkowski Penrose
diagram.}

At this point we have two domain walls separated by a
region of $AdS_{4}$ and outer regions of semi-infinite
Minkowski
space-times.  Clearly, the time-like geodesics
can leave this system and thus another extension
must be specified\refmark{\FOOTIV}.
Depending on the choice of
identifications  there are different possibilities, some
of which we enumerate here and are depicted in Fig. 1
by their conformal diagram.

\REF{\HE}{ S. W. Hawking and G. F. R. Ellis, \sl The Large Scale Structure
of Space-Time. \rm  Cambridge 1973.}

\REF{\FRETAL}{ J.\ Friedmann, M.\ S.\ Morris, I.\ D.\ Novikov,
F.\ Echeverria, G.\ Klinkhammer, K.\ S.\ Thorne, and
U.\ Yurtsever, Phys.\ Rev. {\bf D42}, 1915 (1990).}

(A) One can choose the covering space of the domain wall system, which
makes no identifications and thus contains no closed time-like curves.
The system is an infinite lattice
of semi-infinite Minkowski universes separated by a
continuous $AdS_{4}$ core.
This space-time is simlar to that of the
extreme Reissner-Nordstr\"{o}m (RN) black hole \refmark{\HE}.
However, in the extreme RN space-time there is a time-like
curvature singularity, while here the singularity is replaced by
a domain wall and the curvature is everywhere finite.

(B) One can identify
semi-infinite Minkowski regions living
vertically adjacent to one another.
This identification yields closed time-like curves.
In this case we
see that the closed time-like curves of pure $AdS_{4}$
become a time-machine for the semi-infinite
Minkowski universes: a particle can pass through the wall into the
$AdS_4$ region, cross the Cauchy horizon,  and reemerge
from the wall at an earlier Minkowski time.

(C) One can make an identification
between adjacent Minkowski half-spaces at finite distance from
the wall.
In this case the $AdS_{4}$ space acts as a wormhole in connecting  two
regions of Minkowski space-time\refmark{\FRETAL}.
The Minkowski times at the wormhole mouths(domain walls)
need not be the same, which leads, as in (B),
to the existence of a time-machine for the Minkowski universes.

\REF{\NOTEVI}{As Fig. 1 exhibits, we can extend the coordinates
$u',v'$ across the Cauchy horizon.  Explicitly this is seen by
writing the $1+1$ line element near the horizon as
$ds^{2} = (\alpha z)^{-2}(dt^{2} - dz^{2}) =
[\alpha \sin(1/2(u' - v'))]^{-2} du' dv'$
which has a smooth extension across the
null $u'=\pi, -\pi < v' < \pi$
as well as all the other Cauchy horizons.
The full $3+1$ metric has coordinate
singularites in the $x,y$ directions crossing the null.}

%
%
%

\singlespace

\bf Figure 1  \rm
Conformal diagram of the extended domain wall system.
Coordinates $x,y$ are suppressed; therefore,
each point represents an infinite plane with distances in the
plane conformally compressed by $A(z)$.
The compact null coordinates of $1+1$ Minkowski space-time
define the axes: $u',v' = 2\tan^{-1}[\alpha(t \mp z)]$.
These coordinates can be smoothly extended across the
nulls separating the diamonds\refmark{\NOTEVI}.
Past and future null infinity for the semi-infinite Minkowski regions
are $I-$ and $I+$,  respectively.
The domain walls are the double time-like lines splitting the diamonds.
Time-like geodesics on the Minkowski universes
are arcs beginning at past
time-like infinity and ending at future time-like infinity.
A time-like geodesic on the $AdS_{4}$ side is
the periodic line passing from diamond to diamond.
Cauchy horizons for data placed on the constant time slices
in one diamond are the dashed nulls separating the $AdS_{4}$  patches.
For possibility (B) one makes the identifications
$\beta = \delta$.
For possibility (C) one makes the identifications $\beta = \gamma = \delta$.
Possibility (A) involves no identifications.

\normalspace

\REF{\ROBBERT}{Near the horizon the extremal $3+1$ RN black hole
takes on the Robinson-Bertotti metric $AdS_2\times S^{2}$.
See for example: G. W. Gibbons in: {\it Supersymmetry,
Supergravity and Related Topics}, (F. del Aguila et al. eds.)
World Scientific, Singapore 1985, p. 147;
D. Brill,
Phys.\ Rev.\ {\bf D46}, 1560 (1992);
R. Kallosh and A. Peet, \sl Dilaton Black Holes
near the Horizon \rm , Stanford preprint SU-ITP-92-27, hep-th/9209116.}

One of the most important aspects of these space-times is that they
have Cauchy horizons, where predictability breaks
down at the classical level.
The nulls defining the Cauchy horizons are boundaries beyond which
information placed on an infinite constant time slice in one of the
diamonds is insufficient to specify the evolution
of the data. These nulls are the boundaries
between the diamonds defined by one domain wall, \ie  , for the
first diamond (see Fig. 1) the Cauchy horizon
is at $u' = \pi, -\pi < v' < \pi.$
Additionally, near the horizons,
space-time is
the maximally supersymmetric $AdS_4$ vacuum.
It is interesting to note for motion in the
transverse direction, the
$1+1$ line element
near the Cauchy horizon
can be thought of as the two-dimensional truncation of the
maximally supersymmetric Robinson-Bertotti(RB) metric
$ds^{2}_{RB} = (\rho /GM)^{2}dt^{2} - (GM/\rho)^{2}d\rho^{2},  \    \
\rho\to 0$, where
$z = -\alpha^{-2}\rho^{-1}$ and $\alpha=(GM)^{-1}$.
Recall near the horizon
($r\to GM$)
of the two-dimensional extremal RN black hole,
the metric
$ds^{2}_{RN} = (1 - GM/r)^{2}dt^{2} - (1-GM/r)^{-2}dr^{2}$
is that of RB\refmark{\ROBBERT} where
$\rho =  r - GM$.
Thus, as one passes through the RN horizon at
$r=GM, t=\infty$, the radial motion can
be described by the
metric of $1+1$  $AdS_{2}$.
Outside the RN horizon,  one has
$z^{-1} = -t^{-1} = 0^{-}$
and inside the RN horizon
one uses the next patch $z'^{-1} = -t'^{-1} = 0^{+}$.
In this way the domain wall induces a singularity free
gravitational field with a Cauchy horizon
similar to the horizon of the extremal
RN black hole.

\REF{\FOOTIV}{The possibilities (B) and (C) with their
closed time-like curves
are problematic due to the ambiguity in formulating the Cauchy
problem\refmark{\FRETAL}.
One may speculate with Hawking
that the infinites discussed
in his Chronology Protection Conjecture
[S. W. Hawking, Phys. Rev. \bf D46, \rm 603 (1992)]
could be cancelled due to the usual fermionic and bosonic
cancellations in supersymmetric theories.}

\REF{\ZEROTEMP}{G. W. Gibbons in: {\it Supersymmetry,
Supergravity and Related Topics}, (F. del Aguila et al. eds.)
World Scientific, Singapore 1985, p. 123.}

\REF{\GIBBHULL}{G. W. Gibbons and C. M. Hull, Phys. Lett. \bf 109, \rm
190 (1982).}

We may
examine the thermodynamics of the domain wall system,
in particular,
such properties associated with the covering
space-time, case (A), which has no closed time-like
curves\refmark{\FOOTIV}.
Due to the Cauchy horizon, information will be lost to
the next diamond.
However, there is no Hawking radiation
from this horizon
and therefore the Hawking
temperature of the system is
zero.  This zero temperature result is apparent for the
following reasons:
$(i)$ The wall is  realized as a supersymmetric bosonic
configuration which
is associated with zero temperature\refmark{\ZEROTEMP}.
Recall also that the extreme  RN black hole is a supersymmetric
bosonic configuration\refmark{\GIBBHULL}  and is known to
have zero temperature.
$(ii)$ The Euclidean section of the space-time
does not exhibit a Euclidean
time with finite period.  $(iii)$ The gravitational mass behind the
horizon is zero, which is consistent with its surface gravity,
$ \kappa^{i} = -g_{tt}^{1/2} \Gamma^{\hat{i}}_{\hat{0}\hat{0}}$,
being
zero at the horizon.
Here hats refer to a local orthonormal frame.

\REF{\REVIEW}{For a review see P. C. W. Davies, Rep. Prog. Phys.
\bf 41, \rm 1313 (1978).}

\REF{\SURFACE}{J. David Brown, E. A. Martinez
and J. W. York, Jr., Phys. Rev. Lett. \bf 66, \rm
2281 (1991); F. Wilczek, lecture notes at Princeton
University, 1992, unpublished.}



\REF{\KALLETAL}{R. E. Kallosh, A. D. Linde, T. M. Ort{\' {\i}}n, A. W. Peet,
and A. van Proeyen, \sl Supersymmetry as a Cosmic Censor, \rm  Stanford
preprint SU-ITP-92-13, hep-th/9205027, May 1992.}

The entropy of a black hole can be associated with the
number of states accessible to collapsing matter
forming the hole \refmark{\REVIEW}.  In the case of the
domain wall, the system has zero temperature and thus
only the degeneracy of the
ground state contributes to the
entropy. In this case one has only one state (the solitonic configuration
of the domain wall) and thus the
entropy vanishes.
In other words, there is only one field configuration minimizing
the action and interpolating between the supersymmetric
vacua which produce the horizon and only one smooth space-time
extension across the horizon.
Alternatively, we can
obtain this result by evaluating the ``surface term''
at the horizon\refmark{\SURFACE}.
Using Einstein's equations with the domain wall Ansatz
for the Lagrangian $L$ from Eq. \lagrangian\  yields the on-shell action
$$
I = \int L \sqrt{-g} d^{4}x
  = -{1 \over 2 \kappa} \int {d \over dz}{dA(z) \over dz} dz \int dt dx dy
\eqn\action$$
which is a pure surface term.  It vanishes
for this domain wall system.
A lack of entropy associated with a horizon is not
un-precedented; however, the only known examples are the minimal energy
dilatonic  ``electro-magnetic''
black-holes\refmark{\KALLETAL} which have
space-time singularities.

\REF{\NOTEIII}{Space-time
singularities form when a homogeneous cloud of pressure-free
dust is introduced into pure $AdS_{4}$ due to the focusing
effect of time-like geodesics every $1/2$ the $AdS_{4}$
period: $\pi/2\alpha$: see
F. J. Tipler, C. J. S. Clarke and G. F. R. Ellis in
\sl General Relativity and Gravitation, vol. 2, \rm
edited by A. Held, 1980 Plenum Press, New York.
The same would happen in the
domain wall system if one allowed for an infinite plane of dust
to fall through the wall. However, for realistic finite perturbations
the geometry is stable. In addition, there are
no uncontrollable tidal forces in the background geometry
and thus for physical
particles with repulsive forces (i.e.\ non-zero pressures),
able to withstand the finite tidal forces produced
by themselves and by the
background, no collapse occurs.}

We complete our analysis by examining
the behaviour of classical
objects and
quantum fields on this space-time,
especially near the Cauchy horizons.
Recall for the extreme RN case one has finite
tidal forces and finite quantum field energies
at the horizon; we will see the same results here.
Away from these horizons there is no problem.
Any tidal force or vacuum polarization is due solely
to the domain wall and the constant curvature of the
$AdS_{4}$ space-time.
At the horizon,  the space-time is $AdS_{4}$
which implies  that  if
a classical object can withstand a local tidal force
of magnitude $\alpha^{2}$ in all directions,
passage through the Cauchy horizon into the next diamond
is possible.\refmark{\NOTEIII}


\REF{\BD}{ N. D. Birrell and P. C. W. Davies, \sl Quantum Fields
in Curved Space \rm , chapter 6, Cambridge University Press 1982.}


\REF{\CURVATURES}{
For completeness, we present the calculation of the
renormalized stress-energy tensor
for a massless conformally coupled field on a
conformally flat space-time.  Note we use the
results for a space-time conformal to flat Minkowski
space-time which is appropriate
for one domain wall diamond.
The formulae used can be found in Ref.\refmark{\BD}.
The second order curvature terms are defined by
$S^{\mu}_{\;\;\nu}=
{{1}\over{6}}{ }^{(1)}H^{\mu}_{\;\;\nu}-
{ }^{(3)}H^{\mu}_{\;\;\nu}$
where
${ }^{(1)}H^{\mu}_{\;\;\nu}=
-2R^{,\mu}_{\;\; ;\nu}+2 R^{,\rho}_{\;\; ; \rho}
\delta^{\mu}_{\;\;\nu}-{{1}\over{2}}R^{2}
\delta^{\mu}_{\;\;\nu}+2R R^{\mu}_{\;\;\nu}$
and
${ }^{(3)}H^{\mu}_{\;\;\nu}=
R^{\mu}_{\;\;\rho}R^{\rho}_{\;\;\nu}-{{2}\over{3}}RR^{\mu}_{\;\;\nu}
-{{1}\over{2}}R_{\alpha\beta}R^{\alpha\beta}\delta^{\mu}_{\;\;\nu}
+{{1}\over{4}}R^2 \delta^{\mu}_{\;\;\nu}$.
Defining $H(z)\equiv A'(z)/A(z)$
we find
$A^{2}S^{t}_{\;\;t}  =
 -H'''+{{1}\over{2}}HH''-{{1}\over{4}}(H')^2+H^2H'-{{1}\over{8}}H^4$
and
$A^{2}S^{z}_{\;\;z} =
 -{{3}\over{2}}HH''+{{3}\over{4}}(H')^2+{{3}\over{8}}H^4\;\;$.
By boost invariance along the wall it follows that $S^{t}_{\;\;t}=S^{x}_{\;\;x}
=S^{y}_{\;\;y}$.
}

\REF{\VILENKIN}{A. Vilenkin,
Phys. Lett. \bf 133 B, \rm 177 (1983);
J. Ipser and P. Sikivie, Phys. Rev. \bf D30, \rm 712 (1984).}

In case (A) one can
evaluate the vacuum polarization
which the gravitational field
induces on a massless conformally
coupled quantum field
by
relating it to the local curvature.
A conformally coupled scalar field in a geometry conformally related to the
Minkowski space-time has a vacuum stress-energy tensor equal to \refmark{\BD}
$\langle T^{\mu}_{\;\;\nu}\rangle =-{{\hbar}\over{2880\pi^2}}
S^{\mu}_{\;\;\nu}$
where $S^{\mu}_{\;\;\nu}$
is a second order curvature term
\refmark{\CURVATURES} and we expose factors of
${\hbar}$ to emphasize the quantum nature of this
energy. This stress-energy tensor
is regular everywhere, including the region
near the Cauchy horizon  and it
vanishes  on the Minkowski side.
Note that this is yet
another way of seeing that the Hawking temperature of the
system is zero.
In addition, a major
contribution to the stress-energy tensor is due to the
vacuum polarization  near the wall where the
gradient energy is concentrated.

The classical solution had an exact cancellation of the gravitational mass
of the negative vacuum energy on the $AdS_{4}$ side and the gravitational
mass of the wall. By breaking supersymmetry and letting a scalar field
live on this background one expects the total gravitational mass to change,
and in a semiclassical theory of gravitation this would lead to
gravitational forces in the classically Minkowski region.
To investigate this point, we
calculate the Tolman \refmark{\TOLMAN} mass/area
from the quantum field:
$$
\Sigma(z)_{v}= {{\hbar}\over{2880\pi^2}}\left\{
\left[ {{1}\over{6}}H^3
-H''-HH'\right]^{z}_{-\infty}+{{1}\over{4}}
\int^{z}_{-\infty}\bigl\lbrack 5(H')^2+
\left(H'+H^2\right)^2\bigr\rbrack dz'\right\}\;\; ,
\eqn\vacuummass$$
where $H(z)\equiv A'(z)/A(z)$.
On the $AdS_{4}$ side, the quantum field induces a small
change in the cosmological constant which means the geometry is
virtually unchanged.
Because the boundary term vanishes both at the horizon and in
the Minkowski region far from the wall,
 the gravitational mass of the quantum
vacuum is positive definite.
Hence, as seen from the Minkowski side, there is
a domain wall with
a $positive$ effective gravitational mass near $z=0$.  The gravitational
force on a test particle is thus attractive, in contrast to the
conventional case of $\lambda\phi^4$ non-supersymmetric domain wall which
has repulsive gravitational interactions.
It is well known
\refmark{\VILENKIN} that there exists no static vacuum
solution of Einstein's
field equations describing this space-time. Therefore,
the  quantum corrections
induce time-dependence in the metric.
This example demonstrates the essential role played by supersymmetry in
the existence of static Minkowski-$AdS_4$ domain wall space-times.

\REF{\GIBIII}{G. W.  Gibbons, \sl
Global Structure of Supergravity Domain Wall Space-Times, \rm
DAMTP preprint R-92/38 (in preparation).}

\ack We wish to thank G. W. Gibbons for
correspondence and making us aware
of his related work in progress \refmark{\GIBIII}.
We also benefitted from
discussions with S. Braham, G. Moore, W. Nelson and  A. Peet.
This work was supported in part by
by U.\ S.\ DOE Grant No.\ DOE-EY-76-C-02-3071, SSC Junior Faculty
Fellowship (M. C.),
the NATO Research Grant
No. 900-700 (M.C.),
the Fridtjof Nansen Foundation Grant No.\
152/92 (H. S.), Lise and Arnfinn Heje's Foundation ref. no. 0F0377/1992 (H.
S.),
and by the Norwegian Research Council for Science
and the Humanities(NAVF), Grant No.\ 420.92/022 (H. S.).

\refout
\end